\documentclass[letter]{aa}
\usepackage[varg]{txfonts}
\usepackage{hyperref}
\begin{document}

   \title{Discovery of a dual AGN at $z\simeq 3.3$ with 20 kpc separation\thanks{Based on observations collected at the European Southern Observatory, Paranal, Chile under program ID 094.A-0767(A) (PI: T. Shanks)}}

\subtitle{}
   \author{B. Husemann
          \inst{1}
          \and
          G. Worseck\inst{1,2}
          \and
          F. Arrigoni Battaia\inst{3}
          \and 
          T. Shanks\inst{4}
          }

   \institute{
   Max-Planck-Institut f\"ur Astronomie, K\"onigstuhl 17, D-69117 Heidelberg, Germany, \email{husemann@mpia.de}
   \and
   Institut f\"ur Physik und Astronomie, Universit\"at Potsdam, Karl-Liebknecht-Str. 24/25, D-14476 Potsdam, Germany
   \and
   European Southern Observatory, Karl-Schwarzchild-Str. 2, D-85748 Garching bei M\"unchen, Germany
   \and
   Centre for Extragalactic Astronomy, Durham University, South Road, Durham, DH1 3LE, UK
   }
   \date{Received ...; accepted ...}


\abstract{A prediction of the current paradigm of the hierarchical assembly of galaxies is the presence of supermassive dual black holes at separations of a few kpc or less. In this context, we report the detection of a narrow-line emitter within the extended Ly$\alpha$ nebula ($\sim 120$\,kpc diameter) of the luminous radio-quiet quasi-stellar object (QSO) LBQS~0302$-$0019 at $z=3.286$.  We identify several high-ionization narrow emission lines (\ion{He}{ii}, \ion{C}{iv}, \ion{C}{iii}]) associated with this point-like source, which we have named ``Jil'', which is only $\sim$20\,kpc ($2\farcs9$) away from the QSO in projection. Emission-line diagnostics confirm that the source is likely powered by photoionization of an obscured active galactic nucleus (AGN) three orders of magnitude fainter than the QSO.  The system represents the tightest unobscured/obscured dual AGN currently known at $z>3$, highlighting the power of MUSE to detect these elusive systems.
}


   \maketitle
%

\section{Introduction}

It has long been suggested that the circumgalactic medium (CGM) of QSOs may be detectable in emission via the \ion{H}{i} Ly$\alpha$ line that is powered by recombination radiation, collisional excitation, and Ly$\alpha$ scattering \citep{Rees:1988, Haiman:2001, Cantalupo:2005, Kollmeier:2010}.
Early narrow-band imaging and longslit spectroscopic surveys revealed extended ($\sim 100$\,kpc) Ly$\alpha$ nebulae almost exclusively around radio-loud $2<z<4$ QSOs  \citep[e.g.,][]{Hu:1991, Heckman:1991b}, suggesting an origin in radio jets, as commonly observed in radio galaxies \citep[e.g.,][]{McCarthy:1990,Reuland:2003,Humphrey:2006,Villar-Martin:2007}. 
Subsequent surveys focusing on radio-quiet QSOs found smaller ($\la 70$\,kpc) and fainter ($\sim 10\times$) Ly$\alpha$ nebulae around $\sim 50$\% of the targets \citep{Christensen:2006,North:2012}, but only recent campaigns have ubiquitously detected them and captured their diverse morphologies \citep[e.g.,][]{Hennawi:2013,Borisova:2016,ArrigoniBattaia:2016}. 

Species other than hydrogen enable studies of the ionization conditions and the gas density. Extended \ion{He}{ii}\,$\lambda$1640 and \ion{C}{iv}\,$\lambda$1549 emission is common around radio galaxies and radio-loud QSOs \citep[e.g.,][]{Villar-Martin:2007}, but only $\sim 6$\% of the nebulae around radio-quiet QSOs show these lines \citep{Borisova:2016}. In giant (300--460\,kpc) Ly$\alpha$ nebulae, multiple AGN with separations of several tens of kpc have been discovered via isolated \ion{He}{ii} and metal lines \citep{Cantalupo:2014,Hennawi:2015,Cai:2017,ArrigoniBattaia:2018}.

In this \emph{Letter}, we analyze the environment of the radio-quiet QSO LBQS~0302$-$0019 at $z=3.2859$ \citep{Shen:2016} that has been intensely targeted for studies of the intergalactic medium \citep[IGM, e.g.,][]{Hu:1995} and the impact of foreground galaxies and QSOs on the CGM and IGM \citep[e.g.,][]{Steidel:2003,Jakobsen:2003,Tummuangpak:2014,Schmidt:2017}. In particular, LBQS~0302$-$0019 is one of the few UV-transparent $z>3$ sight lines that allow for \textit{Hubble Space Telescope} UV spectroscopy of intergalactic \ion{He}{ii} Ly$\alpha$ absorption \citep[e.g.,][]{Jakobsen:1994,Syphers:2014}. Here we discuss the detection of various high-ionization lines in its surrounding \ion{H}{i} Ly$\alpha$ nebula, which shows that LBQS~0302$-$0019 is actually an unobscured/obscured dual AGN system with only 20\,kpc projected separation. 

We adopt a flat cosmology with $\Omega_\mathrm{m}=0.3$, $\Omega_\Lambda=0.7$, and $H_0=70$\,km\,s$^{-1}$\,Mpc$^{-1}$. The physical scale at $z=3.286$ is $7.48\,\mathrm{kpc}\,\mathrm{arcsec}^{-1}$.

\section{Observations and results}
\subsection{Observations and data reduction}

\begin{figure*}
 \centering
 \includegraphics[width=0.85\textwidth]{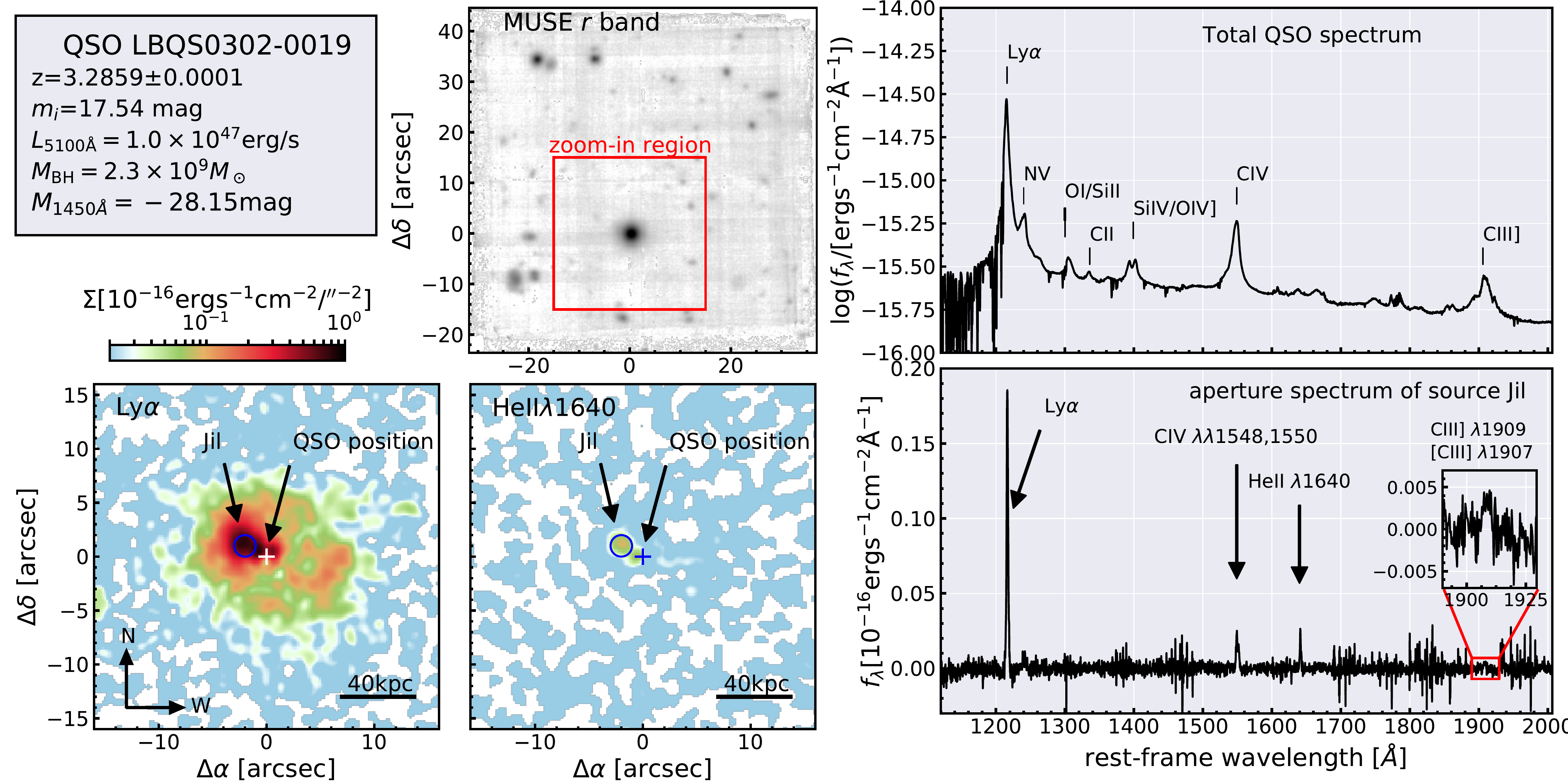}
 \caption{\textit{Top panels:} Basic parameters of LBQS~0302$-$0019 as reported by \citet{Shen:2016} together with an $r$-band image reconstructed from the MUSE data and the QSO spectrum with marked emission lines. The red rectangle on the broad-band image indicates the zoom-in region in the lower panels. \textit{Bottom panels:} Continuum-subtracted narrow-band ($\Delta \lambda=30$\,\AA\ in observed-frame) images of Ly$\alpha$ and \ion{He}{ii}$\lambda1640$ centered on the QSO position (green cross). For visualization, a Gaussian smoothing with a dispersion of one pixel has been applied to suppress the noise. A coadded spectrum from an aperture of 2\arcsec\ diameter centered on the bright emission-line source Jil is shown on the right.}
 \label{fig:overview}
\end{figure*}

Observations of LBQS~0302$-$0019 were taken between October 2014 and January 2015 with the MUSE instrument \citep{Bacon:2010} at the Very Large Telescope. MUSE covers a $\sim1\arcmin\times1\arcmin$ field of view (FoV) with a sampling of 0\farcs2 and spectral coverage from 4750\,\AA\ to 9300\,\AA\ at a spectral resolution of $1800<R<3600$. The observations were split into $11\times 1450$\,s exposures subsequently rotated by 90\degr\ with some small dithering. The median seeing was $\simeq 0\farcs9$. We reduced the data with the latest MUSE data reduction pipeline \citep[v2.0.3.][]{Weilbacher:2012}, which performs all major tasks, i.e.,\ bias subtraction, wavelength calibration, flat-fielding, flux calibration based on photometric standards, and reconstruction of the data cube. While the sky-dominated regions of the FoV are used for an initial sky subtraction, prominent skyline residuals are further suppressed using our own PCA software \citep{Husemann:2016a,Peroux:2017}. The deep reconstructed $r$-band image and the coadded spectrum of LBQS~0302$-$0019 are shown in the top panels of Fig.~\ref{fig:overview}.

\subsection{QSO subtraction and extended Ly$\alpha$ nebula}
To study the extended nebular emission around bright QSOs it is crucial to subtract the point-like QSO emission that  is smeared out due to the seeing, as characterized by the point-spread function (PSF). Various studies have used empirical PSF estimates from the data as a function of wavelength \citep[e.g.,][]{Christensen:2006, Husemann:2014,Herenz:2015, Borisova:2016}. Here we follow the empirical method described in \citet{Borisova:2016}. We constructed a PSF from a median image (150\,\AA\ wide in the observed frame) at each monochromatic slice of the data cube, which is subsequently subtracted after matching the central $0\farcs6\times0\farcs6$.
The subtraction of the QSO reveals a Ly$\alpha$ nebula with a maximum diameter of $\simeq 16\arcsec$ (120\,kpc) as shown in the bottom left panel of Fig.~\ref{fig:overview}.

The Ly$\alpha$ flux integrated over an aperture of 8\arcsec\ radius is $f_{\mathrm{Ly}\alpha}=18.1\times10^{-16}\,\mathrm{erg}\,\mathrm{s}^{-1}\,\mathrm{cm}^{-2}$ which corresponds to a luminosity $L_{\mathrm{Ly}\alpha}=1.7\times 10^{44}\,\mathrm{erg}\,\mathrm{s}^{-1}$. The size and luminosity of this Ly$\alpha$ nebula are similar to those of other radio-quiet QSOs \citep{Borisova:2016}. In this case the Ly$\alpha$ surface brightness distribution is asymmetric, with a bright knot about 2.9\arcsec\ ($\sim$20\,kpc) northeast of the QSO. We  refer to this source as Jil, Klingon for neighbor, with coordinates $\alpha$=03:04:50.03, $\delta$=-00:08:12.5 (J2000), and a peak surface brightness of $\Sigma_{\mathrm{Ly}\alpha}=1.05\times 10^{-16}\,\mathrm{erg}\,\mathrm{s}^{-1}\,\mathrm{cm}^{-2}\,\mathrm{arcsec}^{-2}$.

\subsection{Emission-line diagnostics and photoionization modeling}
\begin{figure*}
\centering
 \includegraphics[width=0.9\textwidth]{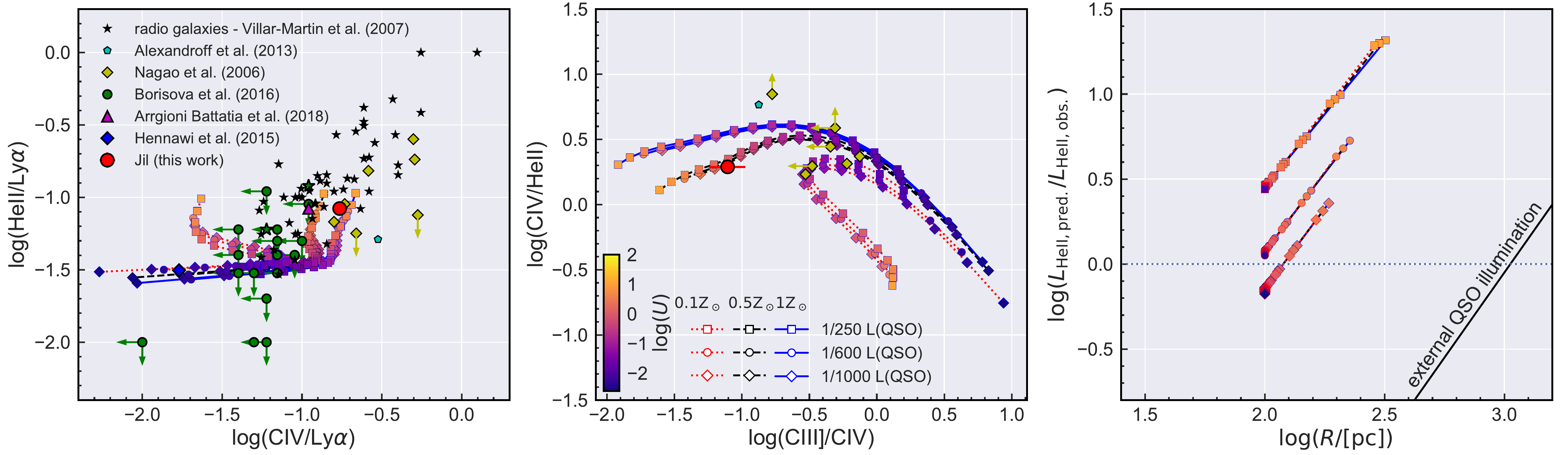}
 \caption{Emission-line diagnostics and photoionization modeling of Jil's spectrum. In all panels we show \texttt{CLOUDY} \citep{Ferland:2013} photoionization model results for an obscured AGN with three different fractions of the LBQS~0302$-$0019 UV luminosity, three metallicities, and a range of ionization parameters $U$ as indicated in the legend and color bar of the middle panel. \textit{Left:} \ion{He}{ii}/Ly$\alpha$ vs. \ion{C}{iv}/Ly$\alpha$ of Jil 20\,kpc  from LBQS~0302$-$0019 (red filled circle) in comparison to various measurements of other Ly$\alpha$ nebulae around QSOs, radio galaxies, and obscured AGN at $z>2$ (see legend).
\textit{Middle:} \ion{C}{iv}/\ion{He}{ii} vs. \ion{C}{iii}]/\ion{C}{iv} diagnostic diagram with symbols as in the left panel. \textit{Right:} Ratio of observed to predicted \ion{He}{ii} luminosity as a function of the radius of a \ion{He}{ii} emitting sphere. Illumination by LBQS~0302$-$0019 results in the black solid line.  }
 \label{fig:line_ratios}
\end{figure*}

The coadded spectrum within a circular aperture of $1\arcsec$ radius around Jil is presented in the bottom right panel of  Fig.~\ref{fig:overview}. We clearly detect \ion{He}{ii} and \ion{C}{iv}$\lambda\lambda 1548,1550$ at $>10\sigma$ significance. Coupling the kinematics to \ion{He}{ii} we also detect [\ion{C}{iii}]\,$\lambda1907$ and \ion{C}{iii}]\,$\lambda1909$ at $3\sigma$ significance. All lines are well fit with single Gaussian profiles whose parameters are listed in Table~\ref{tab:lines}. Ly$\alpha$ is redshifted by 35\,$\mathrm{km}\,\mathrm{s}^{-1}$ (rest frame) compared to the other lines and also shows a significantly larger velocity dispersion  after correcting for the wavelength-dependent spectral resolution of MUSE \citep{Bacon:2017}. Both effects are likely caused by resonant scattering of Ly$\alpha$ photons.

\begin{table}
\begin{small}
 \caption{Emission-line measurements for the source Jil.}\label{tab:lines}
 \begin{tabular}{cccccc}\hline\hline
 Line                           & $f_\mathrm{line}$ & $\log(L_\mathrm{line})$ & $z$\tablefootmark{a} & $\sigma$\\
                                &$\left[10^{-16}\frac{\mathrm{erg}}{\mathrm{s\,cm}^2}\right]$ & $\left[\frac{\mathrm{erg}}{\mathrm{s}}\right]$ && $\left[\frac{\mathrm{km}}{\mathrm{s}}\right]$   \\ \hline
\ion{H}{i} Ly$\alpha$           & $2.15\pm0.04$ & $43.32\pm0.04$ & 3.2887    &  $261\pm7$ \\
 \ion{C}{iv}\,$\lambda1548$     & $0.22\pm0.01$ & $42.33\pm0.05$ & 3.2882    &  $171\pm10$ \\
 \ion{C}{iv}\,$\lambda1550$     & $0.15\pm0.01$ & $42.16\pm0.05$ & 3.2882    &  $171\pm10$ \\
 \ion{He}{ii}\,$\lambda1640$    & $0.18\pm0.01$ & $42.24\pm0.05$ & 3.2882    &  $126\pm10$ \\
 $[$\ion{C}{iii}$]$\,$\lambda1907$      & $0.029\pm0.007$ & $41.45\pm0.13$ & 3.2882    &  $126\pm10$ \\
 \ion{C}{iii}]\,$\lambda1909$   & $0.028\pm0.007$ & $41.43\pm0.13$ & 3.2882    &  $126\pm10$\\\hline
\end{tabular}
\tablefoot{\tablefoottext{a}{Errors on the redshifts are $1\times10^{-4}$.}}
\end{small}
\vspace*{-5mm}
\end{table}

At high redshifts, \ion{He}{ii} has mainly been detected around radio AGN \citep[e.g.,][]{Heckman:1991b, Villar-Martin:2007}, and to date only a few dedicated searches have been performed to detect \ion{He}{ii} and \ion{C}{iv} in the nebulae around radio-quiet QSOs \citep[e.g.,][]{ArrigoniBattaia:2015} and for bright high-redshift galaxies in the re-ionization era \citep[e.g.,][]{Laporte:2017}.
\citet{Borisova:2016} detected \ion{He}{ii} at $2\sigma$ in 1 out of 17 nebulae around radio-quiet QSOs. Isolated \ion{He}{ii} emitters have been found within two of the four known giant Ly$\alpha$ nebulae, 71 and 86\,kpc  from the primary unobscured radio-quiet QSO \citep{Hennawi:2015, ArrigoniBattaia:2018}. The presence of an obscured AGN was invoked in both cases based on the narrow velocity width, the line ratios, and the compactness of the \ion{He}{ii} emitting region. In Fig.~\ref{fig:line_ratios} we show \ion{He}{ii}/Ly$\alpha$ vs. \ion{C}{iv}/Ly$\alpha$ and \ion{C}{iv}/\ion{He}{ii} vs. \ion{C}{iii}]/\ion{C}{iv} for Jil in comparison to various individual nebulae of radio galaxies, unobscured QSOs, and obscured AGN. We also plot the line ratios of a composite spectrum of obscured AGN \citep{Alexandroff:2013}.
The nebular line ratios are inconsistent with the limits for radio-quiet QSOs obtained by \citet{Borisova:2016}, but agree with those of radio-loud QSOs and most obscured AGN. Due to the high \ion{He}{ii} surface brightness, we can derive proper line ratios in a matched aperture.

The detection of several lines allows us to explore ionization properties through a grid of photoionization models with the \texttt{CLOUDY} code \citep[v10.01,][]{Ferland:2013} using the following assumptions and input parameters:
(1) a power-law AGN spectral energy distribution $f_\nu\propto\nu^{\alpha_\nu}$ with $\alpha_\nu=-1.7$ at $\lambda_\mathrm{rest}<912$\,\AA\ \citep{Lusso:2015};
(2) three different ionizing luminosities $L_\mathrm{912\AA}^\mathrm{AGN}=L_\mathrm{912\AA}^\mathrm{QSO}/250$, $L_\mathrm{912\AA}^\mathrm{AGN}=L_\mathrm{912\AA}^\mathrm{QSO}/600$, and $L_\mathrm{912\AA}^\mathrm{AGN}=L_\mathrm{912\AA}^\mathrm{QSO}/1000$, where $L_\mathrm{912\AA}^\mathrm{QSO}$ is 
estimated by scaling the \citet{Lusso:2015} QSO template to the observed SDSS $i$-band magnitude of LBQS~0302$-$0019;
(3) a plane-parallel geometry with an inner distance of 100\,pc from the AGN;
(4) a constant volume number density $n_{\rm H}$ in the range $10^2$--$10^5$\,cm$^{-3}$;
(5) three different metallicities $Z=0.1 Z_{\odot}$, $0.5 Z_{\odot}$, and $1 Z_{\odot}$;
(6) a column density $N_{\rm H}$ determined by the stopping criterion of the calculations at $T=4000$\,K\footnote{For the predictions of interest we found similar results for calculations with a stopping threshold of $T=100$\,K.}.  For each $L_\mathrm{912\AA}^\mathrm{AGN}$, the ionization parameter $U\equiv\Phi_{912\AA}/(cn_{\rm H})$ results from the $n_{\rm H}$ variation, and ranges from $-2.6\lesssim\log U\lesssim2.4$. Our parameter space is similar to works modeling narrow-line regions (NLRs) of obscured AGN \citep[e.g.,][]{Groves:2004, Nagao:2006,Nakajima:2017}.

From the output of the \texttt{CLOUDY} calculations we extract the predictions for the relevant emission-line fluxes and the radius of the \ion{He}{ii} emitting region calculated as the ratio between the column density of \ion{He}{ii} and $n_{\rm H}$. In Fig.~\ref{fig:line_ratios}, we show the predictions of our photoionization models as a function of $U$ for our three AGN luminosities and our three metallicities. We find that an obscured AGN with a luminosity $1000\times$ fainter than the QSO is sufficient to produce the observed \ion{He}{ii} luminosity within an emitting region of $R_{\ion{He}{ii}}<200$\,pc. Our simple models cover the region defined by the observed line ratios, implying $Z<Z_{\odot}$ for the gas around the obscured AGN. At fixed metallicity, models with different $\left(L_\mathrm{912\AA}^\mathrm{AGN}, n_{\rm H}\right)$ yielding the same $U$ parameter are expected to give very similar results (Fig.~\ref{fig:line_ratios}).

\subsection{Intrinsic vs. external AGN ionization source}

Although an obscured AGN appears to be able to power Jil, we also checked whether the QSO can power the emission. We tested this hypothesis by comparing $L_{\ion{He}{ii}}$ with the incident \ion{He}{ii}-ionizing flux 20\,kpc  from the QSO intercepted by a homogeneously filled sphere of radius $R_{\ion{He}{ii}}$. Scaling the \citet{Lusso:2015} broken power-law spectrum to the dereddened SDSS $i$-band magnitude $m_i=17.34$\,mag leads to an extrapolated absolute monochromatic magnitude at the \ion{He}{ii} edge of $M_{228\AA}=-25.78$\,mag. This corresponds to a photon flux of $\Phi(\mathrm{He}^+)=1.85\times10^{10}\,\mathrm{photons}\,\mathrm{s}^{-1}\mathrm{cm}^{-2}$ at a distance of 20\,kpc. Assuming that every emitted \ion{He}{ii} $\lambda$1640 photon requires at least one \ion{He}{ii}-ionizing photon, we can predict the maximum number of emitted \ion{He}{ii} $\lambda$1640 photons from a sphere with radius $R_\ion{He}{ii}$,
$L_{\ion{He}{ii}}(\lambda1640) = \pi R_\ion{He}{ii}^2 \Phi(\mathrm{He}^+)\times h\nu_{1640\AA}\times\frac{\alpha_\mathrm{eff}}{\alpha_\mathrm{B}}$,
where we assumed case B recombination and that all incident ionizing photons passing through the sphere are absorbed. 

The results of this computation are shown in the right panel of Fig.~\ref{fig:line_ratios}, indicating $R_\ion{He}{ii}\gtrapprox$1\,kpc. This size of the \ion{He}{ii} emitting region is a hard lower limit given our simple and very conservative assumption for the QSO ionization scenario. A diameter of $>$2\,kpc would correspond to $>$0\farcs3 projected on the sky. This size is  borderline consistent with the observations at our spatial resolution. A low-luminosity obscured AGN is sufficient to power the observed compact \ion{He}{ii} emission with a much smaller $R_\ion{He}{ii}$, and we do not detect \ion{He}{ii} even coadding the rest of the larger Ly$\alpha$ nebula. We argue that the embedded obscured AGN scenario is much more likely also considering the asymmetry of the nebula. This scenario would naturally explain the Ly$\alpha$ velocity shift due to scattering in the NLR a few 100\,pc away from a highly dust-obscured source, which would not be the case if directly illuminated by the QSO.  Hence, LBQS~0302$-$0019 and Jil form a close dual AGN system with 20\,kpc projected separation.

\section{Discussion}
At low redshifts ($z<1$) numerous dual AGN with kpc-scale separation have been identified through high-resolution X-ray imaging with \textit{Chandra} \citep[e.g.,][]{Koss:2012} or through radio interferometry \citep[e.g.,][]{Fu:2015,MuellerSanchez:2015}. Both methods probe the core emission and are robust in detecting AGN.  However, the sensitivity of \textit{Chandra} is limited and radio-jets can mimic dual AGN signatures in the radio, which makes the methods difficult to apply at high redshifts. Alternatively, the high-ionization [\ion{O}{iii}] $\lambda\lambda 4960,5007$ lines of the NLR have been employed to search for dual AGN. In particular, double-peaked [\ion{O}{iii}] emitters were considered a parent sample for dual AGN candidates \citep[e.g.,][]{Liu:2010a}, but spatially resolved spectroscopy revealed that rotating disks, AGN outflow, jet-cloud interactions are the origin of the double-peaked lines in most cases \citep[e.g.,][]{Fu:2012,Nevin:2016}. The most robust kpc-scale dual AGN systems are always associated with the nuclei of two merging galaxies that are  spatially coincident with AGN signature from the NLR \citep[e.g.,][]{Woo:2014}, from X-rays \citep[e.g.,][]{Liu:2013c,Ellison:2017}, or from radio cores \citep[e.g.,][]{MuellerSanchez:2015}.

Multiple AGN systems at high redshifts are mainly identified as independent bright QSOs in large imaging and spectroscopic surveys \citep[e.g.,][]{Hennawi:2006,Myers:2008}. While most of the known nearby QSOs have separations of several 100\,kpc, a few dual QSOs at $\sim$10\,kpc are identified in the redshift range $0.5<z<2.5$ \citet{Gregg:2002,Pindor:2006,Hennawi:2006,Eftekharzadeh:2017}, and only one QSO pair with $<$20\,kpc separation was reported at $z>3$ by \cite{Hennawi:2010}. Overall the statistics for QSOs at high redshift indicates an excess of QSO clustering at very small separations \citep[e.g.,][]{Hennawi:2006,Myers:2007}. This is somewhat expected as the rapid growth of massive SMBH in the early Universe is related to overdensities as inferred from QSO clustering studies \citep[e.g.,][]{Shen:2007b}. While the prevalence of AGN in major galaxy mergers is highly controversial at low and intermediate redshifts \citep[e.g.,][]{Cisternas:2011,Treister:2012,Villforth:2017}, the role of major mergers for BH growth may be more important at early cosmic times $z>3$.

In case major mergers at high redshifts are more prevalent in triggering AGN, it is possible that many close dual AGN are currently missed because AGN in gas-rich major mergers may often be highly obscured \citep[e.g.,][]{Kocevski:2015,Ricci:2017}.  To detect close dual AGN with at least one obscured companion is challenging at high redshift given the lack of spatial resolution and sensitivity at hard X-rays, and the limited diagnostic power of optical emission lines due to an increasing ionization parameter in high-redshift galaxies \citep[e.g.,][]{Kewley:2013}. Instead, the sensitivity of MUSE allows us to detect the rest-frame far-UV high-ionization emission lines of AGN from which already several obscured AGN at $>$50\,kpc were identified in giant Ly$\alpha$ nebulae around bright QSOs at $z>2$ \citep[e.g.,][]{Hennawi:2015, ArrigoniBattaia:2018} and around the radio-loud QSO PKS~1614$+$051 \citep{Djorgovski:1985,Husband:2015}.  Our dual AGN system is detected with the same method, but at a much smaller projected separation of 20\,kpc, which is already in a regime where the PSF of the bright QSO needs to be subtracted properly for a detection.  In the dual AGN scenario we expect two strongly interacting massive host galaxies to be associated with the two nuclei.  This major merger scenario is testable with deep high-resolution rest-frame optical imaging with \textit{Hubble} and mapping the molecular gas at high angular resolution with ALMA in the sub-mm.

\section{Conclusions}
We report the detection of a \ion{He}{ii} emission-line source, named Jil, at $z=3.28$ that is close to the luminous radio-quiet QSO LBQS~0302-0019. Based on emission-line ratio diagnostics we verified that Jil is ionized most likely by an embedded obscured AGN. With a projected separation of only $\sim$20\,kpc to the QSO, this system represents the tightest unobscured/obscured dual AGN system reported at $z>3$.

High-redshift rest-frame far-UV line diagnostics supersede the classical rest-frame optical line ratios when H$\alpha$ is shifted out of the $K$ band at $z>3$. Furthermore, current X-ray observatories lack the sensitivity and spatial resolution to systematically detect small separation obscured dual AGN at high redshifts. Hence, VLT-MUSE is the ideal instrument to look for tight dual AGN candidates at high-redshift that would be missed otherwise. The ground-layer adaptive optics system of MUSE will further enhance the detectability of these dual AGN, due to a significant increase in spatial resolution and point-source sensitivity.

\bibliography{references}

\begin{thebibliography}{66}
\expandafter\ifx\csname natexlab\endcsname\relax\def\natexlab#1{#1}\fi

\bibitem[{{Alexandroff} {et~al.}(2013){Alexandroff}, {Strauss}, {Greene},
  {Zakamska}, {Ross}, {Brandt}, {Liu}, {Smith}, {Ge}, {Hamann}, {Myers},
  {Petitjean}, {Schneider}, {Yesuf}, \& {York}}]{Alexandroff:2013}
{Alexandroff}, R., {Strauss}, M.~A., {Greene}, J.~E., {et~al.} 2013, \mnras,
  435, 3306

\bibitem[{{Arrigoni Battaia} {et~al.}(2016){Arrigoni Battaia}, {Hennawi},
  {Cantalupo}, \& {Prochaska}}]{ArrigoniBattaia:2016}
{Arrigoni Battaia}, F., {Hennawi}, J.~F., {Cantalupo}, S., \& {Prochaska},
  J.~X. 2016, \apj, 829, 3

\bibitem[{{Arrigoni Battaia} {et~al.}(2015){Arrigoni Battaia}, {Hennawi},
  {Prochaska}, \& {Cantalupo}}]{ArrigoniBattaia:2015}
{Arrigoni Battaia}, F., {Hennawi}, J.~F., {Prochaska}, J.~X., \& {Cantalupo},
  S. 2015, \apj, 809, 163

\bibitem[{{Arrigoni Battaia} {et~al.}(2018){Arrigoni Battaia}, {Prochaska},
  {Hennawi}, {Obreja}, {Buck}, {Cantalupo}, {Dutton}, \&
  {Macci{\`o}}}]{ArrigoniBattaia:2018}
{Arrigoni Battaia}, F., {Prochaska}, J.~X., {Hennawi}, J.~F., {et~al.} 2018,
  \mnras, 473, 3907

\bibitem[{{Bacon} {et~al.}(2010){Bacon}, {Accardo}, {Adjali}, {Anwand},
  {Bauer}, {Biswas}, {Blaizot}, {Boudon}, {Brau-Nogue}, {Brinchmann},
  {Caillier}, {Capoani}, {Carollo}, {Contini}, {Couderc}, {Daguis{\'e}},
  {Deiries}, {Delabre}, {Dreizler}, {Dubois}, {Dupieux}, {Dupuy}, {Emsellem},
  {Fechner}, {Fleischmann}, {Fran{\c c}ois}, {Gallou}, {Gharsa}, {Glindemann},
  {Gojak}, {Guiderdoni}, {Hansali}, {Hahn}, {Jarno}, {Kelz}, {Koehler},
  {Kosmalski}, {Laurent}, {Le Floch}, {Lilly}, {Lizon}, {Loupias}, {Manescau},
  {Monstein}, {Nicklas}, {Olaya}, {Pares}, {Pasquini}, {P{\'e}contal-Rousset},
  {Pell{\'o}}, {Petit}, {Popow}, {Reiss}, {Remillieux}, {Renault}, {Roth},
  {Rupprecht}, {Serre}, {Schaye}, {Soucail}, {Steinmetz}, {Streicher}, {Stuik},
  {Valentin}, {Vernet}, {Weilbacher}, {Wisotzki}, \& {Yerle}}]{Bacon:2010}
{Bacon}, R., {Accardo}, M., {Adjali}, L., {et~al.} 2010, SPIE Conf. Ser., 7735,
  8

\bibitem[{{Bacon} {et~al.}(2017){Bacon}, {Conseil}, {Mary}, {Brinchmann},
  {Shepherd}, {Akhlaghi}, {Weilbacher}, {Piqueras}, {Wisotzki}, {Lagattuta},
  {Epinat}, {Guerou}, {Inami}, {Cantalupo}, {Courbot}, {Contini}, {Richard},
  {Maseda}, {Bouwens}, {Bouch{\'e}}, {Kollatschny}, {Schaye}, {Marino},
  {Pello}, {Herenz}, {Guiderdoni}, \& {Carollo}}]{Bacon:2017}
{Bacon}, R., {Conseil}, S., {Mary}, D., {et~al.} 2017, \aap, 608, A1

\bibitem[{{Borisova} {et~al.}(2016){Borisova}, {Cantalupo}, {Lilly}, {Marino},
  {Gallego}, {Bacon}, {Blaizot}, {Bouch{\'e}}, {Brinchmann}, {Carollo},
  {Caruana}, {Finley}, {Herenz}, {Richard}, {Schaye}, {Straka}, {Turner},
  {Urrutia}, {Verhamme}, \& {Wisotzki}}]{Borisova:2016}
{Borisova}, E., {Cantalupo}, S., {Lilly}, S.~J., {et~al.} 2016, \apj, 831, 39

\bibitem[{{Cantalupo} {et~al.}(2014){Cantalupo}, {Arrigoni-Battaia},
  {Prochaska}, {Hennawi}, \& {Madau}}]{Cantalupo:2014}
{Cantalupo}, S., {Arrigoni-Battaia}, F., {Prochaska}, J.~X., {Hennawi}, J.~F.,
  \& {Madau}, P. 2014, \nat, 506, 63

\bibitem[{{Cantalupo} {et~al.}(2005){Cantalupo}, {Porciani}, {Lilly}, \&
  {Miniati}}]{Cantalupo:2005}
{Cantalupo}, S., {Porciani}, C., {Lilly}, S.~J., \& {Miniati}, F. 2005, \apj,
  628, 61

\bibitem[{{Christensen} {et~al.}(2006){Christensen}, {Jahnke}, {Wisotzki}, \&
  {S{\'a}nchez}}]{Christensen:2006}
{Christensen}, L., {Jahnke}, K., {Wisotzki}, L., \& {S{\'a}nchez}, S.~F. 2006,
  \aap, 459, 717

\bibitem[{{Cisternas} {et~al.}(2011){Cisternas}, {Jahnke}, {Inskip},
  {Kartaltepe}, {Koekemoer}, {Lisker}, {Robaina}, {Scodeggio}, {Sheth},
  {Trump}, {Andrae}, \& {Miyaji}}]{Cisternas:2011}
{Cisternas}, M., {Jahnke}, K., {Inskip}, K.~J., {et~al.} 2011, \apj, 726, 57

\bibitem[{{Djorgovski} {et~al.}(1985){Djorgovski}, {Spinrad}, {McCarthy}, \&
  {Strauss}}]{Djorgovski:1985}
{Djorgovski}, S., {Spinrad}, H., {McCarthy}, P., \& {Strauss}, M.~A. 1985,
  \apjl, 299, L1

\bibitem[{{Eftekharzadeh} {et~al.}(2017){Eftekharzadeh}, {Myers}, {Hennawi},
  {Djorgovski}, {Richards}, {Mahabal}, \& {Graham}}]{Eftekharzadeh:2017}
{Eftekharzadeh}, S., {Myers}, A.~D., {Hennawi}, J.~F., {et~al.} 2017, \mnras,
  468, 77

\bibitem[{{Ellison} {et~al.}(2017){Ellison}, {Secrest}, {Mendel}, {Satyapal},
  \& {Simard}}]{Ellison:2017}
{Ellison}, S.~L., {Secrest}, N.~J., {Mendel}, J.~T., {Satyapal}, S., \&
  {Simard}, L. 2017, \mnras, 470, L49

\bibitem[{{Ferland} {et~al.}(2013){Ferland}, {Porter}, {van Hoof}, {Williams},
  {Abel}, {Lykins}, {Shaw}, {Henney}, \& {Stancil}}]{Ferland:2013}
{Ferland}, G.~J., {Porter}, R.~L., {van Hoof}, P.~A.~M., {et~al.} 2013, \rmxaa,
  49, 137

\bibitem[{{Fu} {et~al.}(2015){Fu}, {Myers}, {Djorgovski}, {Yan}, {Wrobel}, \&
  {Stockton}}]{Fu:2015}
{Fu}, H., {Myers}, A.~D., {Djorgovski}, S.~G., {et~al.} 2015, \apj, 799, 72

\bibitem[{{Fu} {et~al.}(2012){Fu}, {Yan}, {Myers}, {Stockton}, {Djorgovski},
  {Aldering}, \& {Rich}}]{Fu:2012}
{Fu}, H., {Yan}, L., {Myers}, A.~D., {et~al.} 2012, \apj, 745, 67

\bibitem[{{Gregg} {et~al.}(2002){Gregg}, {Becker}, {White}, {Richards},
  {Chaffee}, \& {Fan}}]{Gregg:2002}
{Gregg}, M.~D., {Becker}, R.~H., {White}, R.~L., {et~al.} 2002, \apjl, 573, L85

\bibitem[{{Groves} {et~al.}(2004){Groves}, {Dopita}, \&
  {Sutherland}}]{Groves:2004}
{Groves}, B.~A., {Dopita}, M.~A., \& {Sutherland}, R.~S. 2004, \apjs, 153, 9

\bibitem[{{Haiman} \& {Rees}(2001)}]{Haiman:2001}
{Haiman}, Z. \& {Rees}, M.~J. 2001, \apj, 556, 87

\bibitem[{{Heckman} {et~al.}(1991){Heckman}, {Miley}, {Lehnert}, \& {van
  Breugel}}]{Heckman:1991b}
{Heckman}, T.~M., {Miley}, G.~K., {Lehnert}, M.~D., \& {van Breugel}, W. 1991,
  \apj, 370, 78

\bibitem[{{Hennawi} {et~al.}(2010){Hennawi}, {Myers}, {Shen}, {Strauss},
  {Djorgovski}, {Fan}, {Glikman}, {Mahabal}, {Martin}, {Richards}, {Schneider},
  \& {Shankar}}]{Hennawi:2010}
{Hennawi}, J.~F., {Myers}, A.~D., {Shen}, Y., {et~al.} 2010, \apj, 719, 1672

\bibitem[{{Hennawi} \& {Prochaska}(2013)}]{Hennawi:2013}
{Hennawi}, J.~F. \& {Prochaska}, J.~X. 2013, \apj, 766, 58

\bibitem[{{Hennawi} {et~al.}(2015){Hennawi}, {Prochaska}, {Cantalupo}, \&
  {Arrigoni-Battaia}}]{Hennawi:2015}
{Hennawi}, J.~F., {Prochaska}, J.~X., {Cantalupo}, S., \& {Arrigoni-Battaia},
  F. 2015, Science, 348, 779

\bibitem[{{Hennawi} {et~al.}(2006){Hennawi}, {Strauss}, {Oguri}, {Inada},
  {Richards}, {Pindor}, {Schneider}, {Becker}, {Gregg}, {Hall}, {Johnston},
  {Fan}, {Burles}, {Schlegel}, {Gunn}, {Lupton}, {Bahcall}, {Brunner}, \&
  {Brinkmann}}]{Hennawi:2006}
{Hennawi}, J.~F., {Strauss}, M.~A., {Oguri}, M., {et~al.} 2006, \aj, 131, 1

\bibitem[{{Herenz} {et~al.}(2015){Herenz}, {Wisotzki}, {Roth}, \&
  {Anders}}]{Herenz:2015}
{Herenz}, E.~C., {Wisotzki}, L., {Roth}, M., \& {Anders}, F. 2015, \aap, 576,
  A115

\bibitem[{{Hu} {et~al.}(1995){Hu}, {Kim}, {Cowie}, {Songaila}, \&
  {Rauch}}]{Hu:1995}
{Hu}, E.~M., {Kim}, T.-S., {Cowie}, L.~L., {Songaila}, A., \& {Rauch}, M. 1995,
  \aj, 110, 1526

\bibitem[{{Hu} {et~al.}(1991){Hu}, {Songaila}, {Cowie}, \&
  {Stockton}}]{Hu:1991}
{Hu}, E.~M., {Songaila}, A., {Cowie}, L.~L., \& {Stockton}, A. 1991, \apj, 368,
  28

\bibitem[{{Humphrey} {et~al.}(2006){Humphrey}, {Villar-Mart{\'{\i}}n},
  {Fosbury}, {Vernet}, \& {di Serego Alighieri}}]{Humphrey:2006}
{Humphrey}, A., {Villar-Mart{\'{\i}}n}, M., {Fosbury}, R., {Vernet}, J., \& {di
  Serego Alighieri}, S. 2006, \mnras, 369, 1103

\bibitem[{{Husband} {et~al.}(2015){Husband}, {Bremer}, {Stanway}, \&
  {Lehnert}}]{Husband:2015}
{Husband}, K., {Bremer}, M.~N., {Stanway}, E.~R., \& {Lehnert}, M.~D. 2015,
  \mnras, 452, 2388

\bibitem[{{Husemann} {et~al.}(2016){Husemann}, {Bennert}, {Scharw{\"a}chter},
  {Woo}, \& {Choudhury}}]{Husemann:2016a}
{Husemann}, B., {Bennert}, V.~N., {Scharw{\"a}chter}, J., {Woo}, J.-H., \&
  {Choudhury}, O.~S. 2016, \mnras, 455, 1905

\bibitem[{{Husemann} {et~al.}(2014){Husemann}, {Jahnke}, {S{\'a}nchez},
  {Wisotzki}, {Nugroho}, {Kupko}, \& {Schramm}}]{Husemann:2014}
{Husemann}, B., {Jahnke}, K., {S{\'a}nchez}, S.~F., {et~al.} 2014, \mnras, 443,
  755

\bibitem[{{Jakobsen} {et~al.}(1994){Jakobsen}, {Boksenberg}, {Deharveng},
  {Greenfield}, {Jedrzejewski}, \& {Paresce}}]{Jakobsen:1994}
{Jakobsen}, P., {Boksenberg}, A., {Deharveng}, J.~M., {et~al.} 1994, \nat, 370,
  35

\bibitem[{{Jakobsen} {et~al.}(2003){Jakobsen}, {Jansen}, {Wagner}, \&
  {Reimers}}]{Jakobsen:2003}
{Jakobsen}, P., {Jansen}, R.~A., {Wagner}, S., \& {Reimers}, D. 2003, \aap,
  397, 891

\bibitem[{{Kewley} {et~al.}(2013){Kewley}, {Maier}, {Yabe}, {Ohta}, {Akiyama},
  {Dopita}, \& {Yuan}}]{Kewley:2013}
{Kewley}, L.~J., {Maier}, C., {Yabe}, K., {et~al.} 2013, \apjl, 774, L10

\bibitem[{{Kocevski} {et~al.}(2015){Kocevski}, {Brightman}, {Nandra},
  {Koekemoer}, {Salvato}, {Aird}, {Bell}, {Hsu}, {Kartaltepe}, {Koo}, {Lotz},
  {McIntosh}, {Mozena}, {Rosario}, \& {Trump}}]{Kocevski:2015}
{Kocevski}, D.~D., {Brightman}, M., {Nandra}, K., {et~al.} 2015, \apj, 814, 104

\bibitem[{{Kollmeier} {et~al.}(2010){Kollmeier}, {Zheng}, {Dav{\'e}}, {Gould},
  {Katz}, {Miralda-Escud{\'e}}, \& {Weinberg}}]{Kollmeier:2010}
{Kollmeier}, J.~A., {Zheng}, Z., {Dav{\'e}}, R., {et~al.} 2010, \apj, 708, 1048

\bibitem[{{Koss} {et~al.}(2012){Koss}, {Mushotzky}, {Treister}, {Veilleux},
  {Vasudevan}, \& {Trippe}}]{Koss:2012}
{Koss}, M., {Mushotzky}, R., {Treister}, E., {et~al.} 2012, \apjl, 746, L22

\bibitem[{{Laporte} {et~al.}(2017){Laporte}, {Nakajima}, {Ellis}, {Zitrin},
  {Stark}, {Mainali}, \& {Roberts-Borsani}}]{Laporte:2017}
{Laporte}, N., {Nakajima}, K., {Ellis}, R.~S., {et~al.} 2017, \apj, 851, 40

\bibitem[{{Liu} {et~al.}(2013){Liu}, {Civano}, {Shen}, {Green}, {Greene}, \&
  {Strauss}}]{Liu:2013c}
{Liu}, X., {Civano}, F., {Shen}, Y., {et~al.} 2013, \apj, 762, 110

\bibitem[{{Liu} {et~al.}(2010){Liu}, {Shen}, {Strauss}, \&
  {Greene}}]{Liu:2010a}
{Liu}, X., {Shen}, Y., {Strauss}, M.~A., \& {Greene}, J.~E. 2010, \apj, 708,
  427

\bibitem[{{Lusso} {et~al.}(2015){Lusso}, {Worseck}, {Hennawi}, {Prochaska},
  {Vignali}, {Stern}, \& {O'Meara}}]{Lusso:2015}
{Lusso}, E., {Worseck}, G., {Hennawi}, J.~F., {et~al.} 2015, \mnras, 449, 4204

\bibitem[{{McCarthy} {et~al.}(1990){McCarthy}, {Spinrad}, {Dickinson}, {van
  Breugel}, {Liebert}, {Djorgovski}, \& {Eisenhardt}}]{McCarthy:1990}
{McCarthy}, P.~J., {Spinrad}, H., {Dickinson}, M., {et~al.} 1990, \apj, 365,
  487

\bibitem[{{M{\"u}ller-S{\'a}nchez} {et~al.}(2015){M{\"u}ller-S{\'a}nchez},
  {Comerford}, {Nevin}, {Barrows}, {Cooper}, \& {Greene}}]{MuellerSanchez:2015}
{M{\"u}ller-S{\'a}nchez}, F., {Comerford}, J.~M., {Nevin}, R., {et~al.} 2015,
  \apj, 813, 103

\bibitem[{{Myers} {et~al.}(2007){Myers}, {Brunner}, {Richards}, {Nichol},
  {Schneider}, \& {Bahcall}}]{Myers:2007}
{Myers}, A.~D., {Brunner}, R.~J., {Richards}, G.~T., {et~al.} 2007, \apj, 658,
  99

\bibitem[{{Myers} {et~al.}(2008){Myers}, {Richards}, {Brunner}, {Schneider},
  {Strand}, {Hall}, {Blomquist}, \& {York}}]{Myers:2008}
{Myers}, A.~D., {Richards}, G.~T., {Brunner}, R.~J., {et~al.} 2008, \apj, 678,
  635

\bibitem[{{Nagao} {et~al.}(2006){Nagao}, {Marconi}, \& {Maiolino}}]{Nagao:2006}
{Nagao}, T., {Marconi}, A., \& {Maiolino}, R. 2006, \aap, 447, 157

\bibitem[{{Nakajima} {et~al.}(2017){Nakajima}, {Schaerer}, {Le Fevre},
  {Amorin}, {Talia}, {Lemaux}, {Tasca}, {Vanzella}, {Zamorani}, {Bardelli},
  {Grazian}, {Guaita}, {Hathi}, {Pentericci}, \& {Zucca}}]{Nakajima:2017}
{Nakajima}, K., {Schaerer}, D., {Le Fevre}, O., {et~al.} 2017, ArXiv e-prints

\bibitem[{{Nevin} {et~al.}(2016){Nevin}, {Comerford}, {M{\"u}ller-S{\'a}nchez},
  {Barrows}, \& {Cooper}}]{Nevin:2016}
{Nevin}, R., {Comerford}, J., {M{\"u}ller-S{\'a}nchez}, F., {Barrows}, R., \&
  {Cooper}, M. 2016, \apj, 832, 67

\bibitem[{{North} {et~al.}(2012){North}, {Courbin}, {Eigenbrod}, \&
  {Chelouche}}]{North:2012}
{North}, P.~L., {Courbin}, F., {Eigenbrod}, A., \& {Chelouche}, D. 2012, \aap,
  542, A91

\bibitem[{{P{\'e}roux} {et~al.}(2017){P{\'e}roux}, {Rahmani}, {Quiret},
  {Pettini}, {Kulkarni}, {York}, {Straka}, {Husemann}, \&
  {Milliard}}]{Peroux:2017}
{P{\'e}roux}, C., {Rahmani}, H., {Quiret}, S., {et~al.} 2017, \mnras, 464, 2053

\bibitem[{{Pindor} {et~al.}(2006){Pindor}, {Eisenstein}, {Gregg}, {Becker},
  {Inada}, {Oguri}, {Hall}, {Johnston}, {Richards}, {Schneider}, {Turner},
  {Brasi}, {Hinz}, {Kenworthy}, {Miller}, {Barentine}, {Brewington},
  {Brinkmann}, {Harvanek}, {Kleinman}, {Krzesinski}, {Long}, {Neilsen},
  {Newman}, {Nitta}, {Snedden}, \& {York}}]{Pindor:2006}
{Pindor}, B., {Eisenstein}, D.~J., {Gregg}, M.~D., {et~al.} 2006, \aj, 131, 41

\bibitem[{{Rees}(1988)}]{Rees:1988}
{Rees}, M.~J. 1988, \nat, 333, 523

\bibitem[{{Reuland} {et~al.}(2003){Reuland}, {van Breugel}, {R{\"o}ttgering},
  {de Vries}, {Stanford}, {Dey}, {Lacy}, {Bland-Hawthorn}, {Dopita}, \&
  {Miley}}]{Reuland:2003}
{Reuland}, M., {van Breugel}, W., {R{\"o}ttgering}, H., {et~al.} 2003, \apj,
  592, 755

\bibitem[{{Ricci} {et~al.}(2017){Ricci}, {Bauer}, {Treister}, {Schawinski},
  {Privon}, {Blecha}, {Arevalo}, {Armus}, {Harrison}, {Ho}, {Iwasawa},
  {Sanders}, \& {Stern}}]{Ricci:2017}
{Ricci}, C., {Bauer}, F.~E., {Treister}, E., {et~al.} 2017, \mnras, 468, 1273

\bibitem[{{Schmidt} {et~al.}(2017){Schmidt}, {Worseck}, {Hennawi}, {Prochaska},
  \& {Crighton}}]{Schmidt:2017}
{Schmidt}, T.~M., {Worseck}, G., {Hennawi}, J.~F., {Prochaska}, J.~X., \&
  {Crighton}, N.~H.~M. 2017, \apj, 847, 81

\bibitem[{{Shen}(2016)}]{Shen:2016}
{Shen}, Y. 2016, \apj, 817, 55

\bibitem[{{Shen} {et~al.}(2007){Shen}, {Strauss}, {Oguri}, {Hennawi}, {Fan},
  {Richards}, {Hall}, {Gunn}, {Schneider}, {Szalay}, {Thakar}, {Vanden Berk},
  {Anderson}, {Bahcall}, {Connolly}, \& {Knapp}}]{Shen:2007b}
{Shen}, Y., {Strauss}, M.~A., {Oguri}, M., {et~al.} 2007, \aj, 133, 2222

\bibitem[{{Steidel} {et~al.}(2003){Steidel}, {Adelberger}, {Shapley},
  {Pettini}, {Dickinson}, \& {Giavalisco}}]{Steidel:2003}
{Steidel}, C.~C., {Adelberger}, K.~L., {Shapley}, A.~E., {et~al.} 2003, \apj,
  592, 728

\bibitem[{{Syphers} \& {Shull}(2014)}]{Syphers:2014}
{Syphers}, D. \& {Shull}, J.~M. 2014, \apj, 784, 42

\bibitem[{{Treister} {et~al.}(2012){Treister}, {Schawinski}, {Urry}, \&
  {Simmons}}]{Treister:2012}
{Treister}, E., {Schawinski}, K., {Urry}, C.~M., \& {Simmons}, B.~D. 2012,
  \apjl, 758, L39

\bibitem[{{Tummuangpak} {et~al.}(2014){Tummuangpak}, {Bielby}, {Shanks},
  {Theuns}, {Crighton}, {Francke}, \& {Infante}}]{Tummuangpak:2014}
{Tummuangpak}, P., {Bielby}, R.~M., {Shanks}, T., {et~al.} 2014, \mnras, 442,
  2094

\bibitem[{{Villar-Mart{\'{\i}}n} {et~al.}(2007){Villar-Mart{\'{\i}}n},
  {S{\'a}nchez}, {Humphrey}, {Dijkstra}, {di Serego Alighieri}, {De Breuck}, \&
  {Gonz{\'a}lez Delgado}}]{Villar-Martin:2007}
{Villar-Mart{\'{\i}}n}, M., {S{\'a}nchez}, S.~F., {Humphrey}, A., {et~al.}
  2007, \mnras, 378, 416

\bibitem[{{Villforth} {et~al.}(2017){Villforth}, {Hamilton}, {Pawlik},
  {Hewlett}, {Rowlands}, {Herbst}, {Shankar}, {Fontana}, {Hamann}, {Koekemoer},
  {Pforr}, {Trump}, \& {Wuyts}}]{Villforth:2017}
{Villforth}, C., {Hamilton}, T., {Pawlik}, M.~M., {et~al.} 2017, \mnras, 466,
  812

\bibitem[{{Weilbacher} {et~al.}(2012){Weilbacher}, {Streicher}, {Urrutia},
  {Jarno}, {P{\'e}contal-Rousset}, {Bacon}, \& {B{\"o}hm}}]{Weilbacher:2012}
{Weilbacher}, P.~M., {Streicher}, O., {Urrutia}, T., {et~al.} 2012, SPIE Conf.
  Ser., 8451

\bibitem[{{Woo} {et~al.}(2014){Woo}, {Cho}, {Husemann}, {Komossa}, {Park}, \&
  {Bennert}}]{Woo:2014}
{Woo}, J.-H., {Cho}, H., {Husemann}, B., {et~al.} 2014, \mnras, 437, 32

\end{thebibliography}

\end{document}